\documentclass[11pt]{article}
\usepackage[margin=1in]{geometry}
\usepackage[T1]{fontenc}
\usepackage{mathptmx}
\usepackage[numbers,sort&compress]{natbib}
\usepackage{graphicx}
\usepackage{booktabs}
\usepackage{array}
\newcolumntype{L}[1]{>{\raggedright\arraybackslash}p{#1}}
\usepackage{multirow}
\usepackage{amsmath}
\usepackage{amssymb}
\usepackage{xcolor}
\usepackage{tikz}
\usetikzlibrary{positioning,fit,arrows.meta}
\usepackage[hidelinks]{hyperref}
\usepackage{caption}
\newcommand{\sd}[1]{\textsubscript{$\pm$#1}}

\title{From Sound to Occupancy:\\ Privacy-Aware Audio Sensing for Buildings}
\author{Myeongwon Chae\\
School of Architecture\\
University of Illinois Urbana-Champaign\\
\texttt{mc130@illinois.edu}
\and
Seonghyuk Son\thanks{Corresponding author.}\\
School of Architecture\\
University of Kentucky\\
\texttt{sh.son@uky.edu}}
\date{}

\begin{document}
\maketitle

\begin{abstract}
Occupancy-based controls regulate lighting and HVAC according to whether a space is in use. Passive infrared (PIR) sensors, the most common choice, respond to motion and can report an occupied room as vacant when its occupants remain still. Sound offers an alternative cue, since seated occupants type, turn pages and cough, but microphones raise privacy concerns. We therefore study a two-stage design: an on-device encoder maps each second of audio to a short embedding, and a server-side temporal model turns the stream of embeddings into an occupancy state. To our knowledge, no publicly available dataset pairs office audio with occupancy labels, so we evaluate the two stages separately on public data. The encoder is trained on synthetic scenes that mix empty-room recordings, room impulse responses, office sounds and read speech, and tested on unseen rooms, recordings and speakers for whether a second contains a sound made by a person. On 11 real office recordings (DCASE 2013) it reaches an AUC of 0.87, against 0.76 for log-mel statistics. Attacks on the transmitted embeddings recover at most 1\% of the spoken words, against 94\% from log-mel frames, but still identify the speaker among 40 in 69 to 77\% of the utterances. Noise-floor normalization with adversarial training, an 8-dimensional embedding, or sending only class probabilities lowers this to 38 to 55\% without hurting detection. On HPDmobile, which pairs home occupancy labels with an amplitude envelope of the audio, a causal temporal model pretrained with a hierarchical contrastive loss matches hand-crafted features (AUC 0.75) and does better when only one labelled day per home is available. These results make the two-stage design a practical basis for privacy-aware acoustic occupancy sensing, with speaker identity as the one leak still to close.
\end{abstract}

\section{Introduction}

Occupancy sensors are widely used to control lighting and heating, ventilation and air conditioning (HVAC) in buildings, and the energy they save depends on how reliably they detect people \citep{guo2010lighting}. The most common of them, passive infrared (PIR) sensors, respond to motion \citep{shokrollahi2024pir}, but much of office work involves little of it: people read, type, think or listen for long stretches while barely moving. When an occupant stays still for longer than the sensor's time delay, the room is reported vacant and its lights or ventilation are turned down. Extending the timeout can reduce premature switch-offs during quiet occupancy, but it also delays the shutdown of lighting or ventilation after occupants leave \citep{guo2010lighting}. This limitation motivates investigating additional cues of presence during periods of little movement.

Other sensors have been tried: active ultrasonic systems infer occupancy from how a room answers emitted chirps \citep{shih2015chirps,shih2016aures}, and multimodal systems add cameras, microphones and environmental sensors \citep{jacoby2021hpdmobile,tan2022fusion}. Microphones can capture sounds produced during otherwise sedentary work, such as typing, mouse clicks, coughing, chair movements and page turns. Such sounds have been used to detect occupancy and activity in commercial and residential buildings \citep{ghaffarzadegan2017occupancy,valle2016abroa,santiago2021audio}. Microphones also capture conversations, however, so a parallel line of work uses audio features designed to limit speech-content recovery while supporting speech or activity detection \citep{wyatt2007conversation,parthasarathi2009privacy,parthasarathi2011privacy} and occupancy estimation \citep{alhossain2023crowdotic,williams2023privacy}.

Three problems remain before these approaches can control lighting and ventilation in offices. First, discarding the words does not make audio private. Common spectral features can be turned back into intelligible speech \citep{milner2002reconstruction,shao2005f0}, and where attacks have been tested the speaker remained exposed: age and gender could be inferred after speech was masked \citep{chhaglani2025featuresense}, and individual speakers identified from perturbed audio that hid the words \citep{williams2023privacy}. What an attacker can recover from the data a sensor sends, both the words and the speaker, therefore has to be measured. Second, occupancy is a state with memory. A person working quietly may make a sound only every few minutes. A detector that judges each short window on its own cannot tell such a room from an empty one. Third, to our knowledge no public dataset pairs office audio with occupancy labels. The datasets with occupancy labels come from homes: SINS and its DCASE 2018 subset label daily activities in raw audio, including absence \citep{dekkers2017sins,dekkers2018dcase}, and HPDmobile provides long-term labels but only a processed amplitude envelope \citep{jacoby2021hpdmobile}. Office recordings, such as those of DCASE 2013, are labelled with sound events rather than presence \citep{stowell2015dcase}.

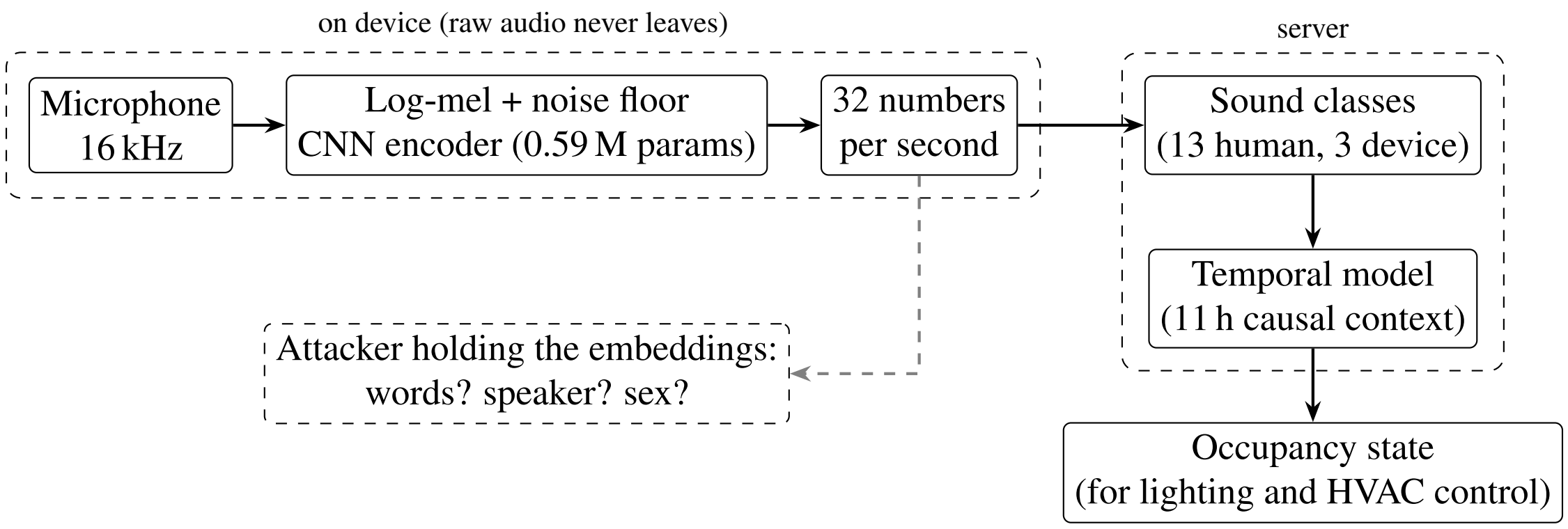
\begin{figure}[t]
\centering
\begin{tikzpicture}[
  box/.style={draw, rounded corners=2pt, align=center, font=\small, minimum height=9mm, inner sep=3pt},
  arr/.style={-{Stealth[length=2mm]}, thick},
  lab/.style={font=\scriptsize, align=center}]
\node[box] (mic) {Microphone\\16\,kHz};
\node[box, right=5mm of mic] (enc) {Log-mel + noise floor\\CNN encoder (0.59\,M params)};
\node[box, right=5mm of enc] (emb) {32 numbers\\per second};
\node[box, right=12mm of emb] (head) {Sound classes\\(13 human, 3 device)};
\node[box, below=7mm of head] (temp) {Temporal model\\(11\,h causal context)};
\node[box, below=7mm of temp] (ctrl) {Occupancy state\\(for lighting and HVAC control)};
\node[draw, dashed, rounded corners=4pt, fit=(mic)(enc)(emb), inner sep=6pt, label={[lab]above:on device (raw audio never leaves)}] {};
\node[draw, dashed, rounded corners=4pt, fit=(head)(temp), inner sep=6pt, label={[lab]above:server}] {};
\node[box, dashed, below=14mm of enc] (att) {Attacker holding the embeddings:\\words? speaker? sex?};
\draw[arr] (mic) -- (enc);
\draw[arr] (enc) -- (emb);
\draw[arr] (emb) -- (head);
\draw[arr] (head) -- (temp);
\draw[arr] (temp) -- (ctrl);
\draw[arr, dashed, gray] (emb.south) |- (att.east);
\end{tikzpicture}
\caption{Two-stage design. The device sends one embedding per second; the server detects sound classes and infers an occupancy state from hours of context. The dashed box is the attacker, who holds what the device transmits. The two stages are evaluated separately in this study.}
\label{fig:overview}
\end{figure}

This paper addresses the three problems with public data only, and its main contribution is an evaluation rather than a new architecture: for each audio representation we measure both how well it detects human-generated sounds and what an attacker can recover from it. Figure~\ref{fig:overview} shows the two-stage sensor we evaluate: a small encoder on the device turns each second of audio into a short vector, and a temporal model on the server reads the stream of vectors and tracks whether the room is occupied. Without office occupancy labels the two stages are evaluated separately. The encoder is trained on synthetic scenes built from public recordings of empty rooms, room impulse responses, isolated office sounds and read speech, and tested on scenes from unseen rooms and on real office recordings for human-generated sounds, that is, whether a second contains a sound made by a person, the evidence the temporal model accumulates. Every representation is then attacked with speech reconstruction followed by recognition, speaker identification and sex classification. The temporal model is pretrained on unlabelled sensor streams with the hierarchical contrastive loss of TS2Vec \citep{yue2022ts2vec} and evaluated on presence in the homes of HPDmobile.

\section{Related Work}

\paragraph{Acoustic occupancy and activity sensing.} Occupancy has been detected from audio in commercial and residential environments \citep{ghaffarzadegan2017occupancy}, with Gaussian mixtures and hidden Markov models \citep{valle2016abroa}, and with engineered audio features \citep{santiago2021audio}. Williams et al.\ counted overlapping talkers (1 to 10) in simulated cocktail-party mixtures from randomly perturbed audio frames \citep{williams2023privacy}. Crowdotic estimated crowd density in hospital waiting rooms from non-speech audio \citep{alhossain2023crowdotic}. Ambient audio has also been proposed for estimating indoor airborne transmission risk factors, including how occupants are distributed \citep{chhaglani2025ambient}.

\paragraph{Privacy of audio representations.} Chhaglani et al.\ described the privacy risks of audio classification and proposed privacy-preserving features \citep{chhaglani2024privacy}. Our attacks follow the approach of Williams et al.\ \citep{williams2023privacy}, but because our representations are not audio, we first reconstruct speech from them before recognition. Adversarial training with a gradient reversal layer \citep{ganin2015grl} is a common way to remove an unwanted attribute from a representation, but the attribute often remains recoverable by a classifier trained afterwards, for demographic attributes in text \citep{elazar2018adversarial} as for speaker identity in speech representations \citep{srivastava2019privacy}. At the hardware level, PrivacyMic provides settings that exclude speech frequencies or the entire audible range \citep{iravantchi2021privacymic}. Our approach processes audible-band audio locally and limits the information transmitted from the device. Privacy concerns in building data have also motivated learning from anonymized data \citep{choi2026pptl}.

\paragraph{Self-supervised representations.} Contrastive learning yields general-purpose audio representations \citep{saeed2020cola}. TS2Vec learns timestamp-level representations of time series with a hierarchical contrastive loss over random overlapping crops \citep{yue2022ts2vec}. We use this loss to pretrain the temporal occupancy model on unlabelled sensor streams.

\section{Problem Setup and Threat Model}
\label{sec:setup-threat}

A sensing device in a room records audio at 16\,kHz. For each second $t$ it computes a representation $v_t$ and sends it to a server; raw audio is neither stored nor transmitted. In most experiments $v_t$ is an embedding; in one variant it is the vector of class probabilities computed on the device. The server outputs a human-sound score $s_t$ per second and an occupancy state $y_t \in \{0,1\}$ for control. For utility, the encoder should detect human-generated sounds, including subtle sounds associated with sedentary activities, in rooms not used during training. For privacy, the transmitted representation $v_t$ should minimize the recoverability of spoken content and speaker identity. For deployability, the encoder must be small enough to run on the device.

Human-generated sound detection and occupancy inference are distinct tasks. The former determines whether a one-second segment contains a sound produced by a person, whereas the latter determines whether someone is present, even when no sound is produced. Our office experiments evaluate human-generated sound detection, whereas the residential experiments evaluate home-level occupancy inference using audio-envelope data.

The adversary obtains the transmitted representations, for example through a compromised server, and knows the encoder. In the strongest attack setting, the attacker also has access to labelled, clean close-talking recordings of the target speakers. It tries to (a) reconstruct and transcribe speech, (b) identify the speaker among a known set, and (c) infer the speaker's sex. Attacks on the device itself, which has access to raw audio, are out of scope. We give no formal privacy guarantee and report only what these attacks recover.

\section{Method}
\label{sec:method}

\paragraph{On-device encoder.} Each second of audio is converted to 64-band log-mel energies $m_{f,\tau}$ for band $f$ and frame $\tau$ (25\,ms windows, 10\,ms hop), giving a $64\times100$ patch. Noise-floor normalization (NF) subtracts, in each band, the 10th percentile over the set $\mathcal{W}$ of the 1,000 frames in the 10-s block that contains the current second:
\begin{equation}
\tilde m_{f,\tau} = m_{f,\tau} - Q_{0.1}\big(\{m_{f,\tau'} : \tau' \in \mathcal{W}\}\big).
\label{eq:nf}
\end{equation}
The features then describe deviations from the background of the room and microphone. After batch normalization of each band, a convolutional network $g_\phi$ maps each normalized one-second patch $\tilde M_t$ to an embedding $z_t = g_\phi(\tilde M_t) \in \mathbb{R}^d$. It has four blocks of two $3\times3$ convolutions (32, 64, 128 and 128 channels) with batch normalization and ReLU, and $2\times2$ average pooling between blocks. Its output is averaged over frequency, mean- and max-pooled over time, and projected to $d$ dimensions, with $d=32$ unless stated otherwise. The network has 0.59 million parameters and processes each one-second audio segment independently. Through Eq.~\eqref{eq:nf}, however, the features also depend on the 10-s block, so the device processes audio in 10-s blocks and emits the embeddings of a block once the block is complete, up to 10\,s after the sound. When sound fills the whole block, the 10th percentile is no longer the level of the empty room; we did not treat this case separately.

\paragraph{Activity head.} On the server, a layer-normalized two-layer perceptron $h_\psi$ with 128 hidden units predicts the probabilities of 16 sound classes, $p_t = \sigma(h_\psi(z_t)) \in [0,1]^{16}$. Thirteen classes are categorized as human-generated sounds (clear throat, cough, door slam, drawer, keyboard, keys, knock, laughter, mouse click, page turn, pen drop, speech, light switch), while the remaining three are categorized as device-generated sounds (alert, phone ringing, printer). The human-sound score of a second is the largest probability among the human classes $\mathcal{H}$,
\begin{equation}
s_t = \max_{c \in \mathcal{H}} p_{t,c}.
\label{eq:score}
\end{equation}
Encoder and head are trained with the binary cross-entropy over the $T=10$ seconds and 16 classes of a scene,
\begin{equation}
\mathcal{L}_{\mathrm{ev}} = -\frac{1}{16T} \sum_{t=1}^{T}\sum_{c=1}^{16} \Big[ y_{t,c}\log p_{t,c} + (1-y_{t,c})\log(1-p_{t,c}) \Big],
\label{eq:bce}
\end{equation}
where $y_{t,c}=1$ if class $c$ is active in second $t$. The encoder and head classify each second on its own. Unlike sound event detectors, which use context across frames and seconds \citep{cakir2017crnn}, they have no temporal context; that is left to the occupancy model.

\paragraph{Speaker-adversarial training.} An adversary $a_\omega$, a perceptron with two hidden layers of 256 units, predicts the speaker of a scene among 251 training speakers. Its input is the mean $\bar z$ and the element-wise maximum $\hat z$ of the embeddings over the speech-active seconds, and its loss for the true speaker $k$ is $\mathcal{L}_{\mathrm{adv}} = -\log a_\omega(k \mid \bar z, \hat z)$. Scenes without speech (60\% of them on average) contribute only to $\mathcal{L}_{\mathrm{ev}}$. A gradient reversal layer \citep{ganin2015grl} between encoder and adversary leaves the forward pass unchanged and multiplies the gradient by $-\lambda_p$ on the way back, so that training solves
\begin{equation}
\min_{\omega}\ \mathcal{L}_{\mathrm{adv}}, \qquad \min_{\phi,\psi}\ \mathcal{L}_{\mathrm{ev}} - \lambda_p\,\mathcal{L}_{\mathrm{adv}}.
\label{eq:adv}
\end{equation}
The adversary learns to identify the speaker, while the encoder keeps what the sound classes need and removes what helps the adversary. Following \citet{ganin2015grl}, the weight grows with training progress $p \in [0,1]$ as
\begin{equation}
\lambda_p = \lambda \left( \frac{2}{1+e^{-10p}} - 1 \right), \qquad \lambda \in \{1, 3\}.
\label{eq:lambda}
\end{equation}

\paragraph{Temporal occupancy model.} For long-term occupancy we use a causal dilated convolutional encoder $f_\theta$ that maps the sequence of input vectors up to time $t$ to a representation $r_t \in \mathbb{R}^{64}$. The inputs are what the device transmits: in a deployment, the embeddings $z_t$ of the on-device encoder; in our HPDmobile experiments, the per-slot statistics of the audio envelope described in Section~\ref{sec:hpd-protocol}, since that dataset releases no raw audio. Its receptive field is long because, in the homes of Section~\ref{sec:homes}, sound concentrates around arrivals and departures and the occupancy state has to be carried through long quiet stretches. A linear input projection to 64 channels is followed by 10 residual blocks and a $1\times1$ output convolution to 64 channels. Each block has two convolutions with 64 channels, kernel size 3, dilation $2^i$ ($i = 0, \dots, 9$) and GELU activations, padded on the past side only. The receptive field is 4,093 steps, or 11.4 hours at one step per 10\,s.

We pretrain $f_\theta$ without labels using the hierarchical contrastive loss of TS2Vec \citep{yue2022ts2vec}. Two overlapping random crops of each sequence are encoded with random timestamp masking, which zeroes each projected input step with probability 0.5, giving representations $r_{i,t}$ and $r'_{i,t}$ of sequence $i$ at each timestamp $t$ in the overlap. The temporal loss contrasts each timestamp with the other timestamps $\Omega$ of the same sequence. The instance-wise loss contrasts it with the same timestamp of the other sequences in a batch of size $B$:
\begin{align}
\ell^{\mathrm{temp}}_{i,t} &= -\log \frac{\exp(r_{i,t}\cdot r'_{i,t})}{\sum_{t'\in\Omega}\big(\exp(r_{i,t}\cdot r'_{i,t'}) + \mathbf{1}[t\neq t']\,\exp(r_{i,t}\cdot r_{i,t'})\big)},
\label{eq:temp}\\
\ell^{\mathrm{inst}}_{i,t} &= -\log \frac{\exp(r_{i,t}\cdot r'_{i,t})}{\sum_{j=1}^{B}\big(\exp(r_{i,t}\cdot r'_{j,t}) + \mathbf{1}[i\neq j]\,\exp(r_{i,t}\cdot r_{j,t})\big)}.
\label{eq:inst}
\end{align}
Both losses are averaged over sequences, timestamps and the two crops and combined with equal weight. The representations are then max-pooled along time by a factor of 2, and the loss is computed again. This repeats until a single timestamp remains, where only the instance-wise term applies. The pretraining loss is the mean over these levels.

For prediction we fit a class-balanced logistic regression on the frozen representation and the time of day (linear probe),
\begin{equation}
P(y_t = 1) = \sigma\big(w^\top [\, r_t;\ \sin 2\pi\tau_t;\ \cos 2\pi\tau_t \,] + b\big),
\label{eq:probe}
\end{equation}
where $\tau_t \in [0,1)$ is the time of day as a fraction of 24 hours. We also report fine-tuning the pretrained network and training the same network from scratch; in both cases a linear layer on $[\,r_t;\ \sin 2\pi\tau_t;\ \cos 2\pi\tau_t\,]$ is trained together with $f_\theta$.

Pretraining draws training windows, contiguous stretches of 3,000 steps (8.3\,h at 10\,s per step), from the pretraining homes, skipping windows in which more than half of the steps have no data, and runs 600 iterations of AdamW (learning rate $10^{-3}$) on batches of 8 windows; the common part of the two crops has a random length of 2 to 3,000 steps. Fine-tuning and training from scratch use binary cross-entropy with the positive class weighted by the ratio of negative to positive training labels, for 800 iterations of AdamW (learning rate $10^{-3}$, weight decay $10^{-2}$) on batches of 8 windows of 6,144 steps (17\,h at 10\,s per step), with no loss on the first third of each window, where the causal context is short. The probe is an $L_2$-regularized logistic regression (regularization strength $C=1$, class weights inversely proportional to class frequency) on standardized inputs, fitted on every sixth labelled step, that is, one step per minute.

\section{Experimental Setup}
\label{sec:setup}

\subsection{Data}
Table~\ref{tab:data} lists the public datasets. We used only recordings whose licenses allow research use, and we report the license stated by each host.

\paragraph{HPDmobile.} Jacoby et al.\ instrumented six homes in Boulder, Colorado, apartments and single-family houses with varying numbers of occupants, for four to eight weeks each in 2019 \citep{jacoby2021hpdmobile}. Each home had four to six sensor hubs in common areas, never in bedrooms or bathrooms, and each hub recorded monochrome images at one frame per second and 10-s audio clips at 8\,kHz. Raw audio was not released: every clip was mean-shifted, full-wave rectified and downsampled by a factor of 100 before publication, which removes intelligible speech. Occupancy ground truth combines phone geofencing logs with paper sign-in sheets, reconciled by the authors and corrected where the camera showed a person during a period marked vacant, and is given as a home-level occupied or vacant state for every 10-s interval. Audio from three to five hubs per home was available to us.

\paragraph{DCASE 2013 office sounds.} The event detection task of the first DCASE challenge \citep{stowell2015dcase} covers 16 office sound classes. Its training material is a set of isolated recordings of each class, and its office live (OL) test set consists of 11 recordings of one to three minutes in several office environments, in which people performed scripted, non-overlapping sequences of the 16 sounds; the recordings were annotated by two annotators.

\paragraph{DCASE 2016 Task 2.} This task reused 11 of the 16 DCASE 2013 classes, dropping alert, mouse click, pen drop, printer and light switch as too short or too broad, with new isolated sounds recorded in offices at LS2N in France \citep{lafay2017dcase2016}. The release contains 20 training sounds per class and a public test set of 54 two-minute mixtures that the organizers synthesized from a pool of 40 sounds per class at event-to-background ratios of $-6$, 0 and $+6$\,dB, with monophonic and polyphonic conditions.

\paragraph{LibriSpeech.} LibriSpeech is a corpus of read English speech derived from public-domain audiobooks, sampled at 16\,kHz and segmented into utterances with transcripts and speaker labels \citep{panayotov2015librispeech}. The train-clean-100, dev-clean and test-clean subsets contain 251, 40 and 40 speakers, with no speaker shared between subsets; the "clean" subsets hold the speakers whose recordings a baseline recognizer transcribed most accurately, and the recordings are close-talking and free of noticeable noise.

\paragraph{BUT ReverbDB.} The BUT Speech@FIT Reverb Database was recorded by Brno University of Technology in nine rooms: three offices, two lecture rooms, a meeting room, a conference room, a hotel room and a stairwell \citep{szoke2019but}. Each room was measured with 31 microphones and two to ten loudspeaker positions, giving room impulse responses from exponential sine sweeps and, for every loudspeaker position, a recording of the empty room.

\begin{table}[t]
\centering\small
\caption{Summary of public datasets used for acoustic event detection, privacy evaluation, synthetic scene generation, and temporal occupancy inference.}
\label{tab:data}
\begin{tabular}{@{}L{2.6cm}L{5.2cm}L{2.6cm}L{1.6cm}L{2.4cm}@{}}
\toprule
Dataset & Content used & Labels & License & Role \\
\midrule
HPDmobile \citep{jacoby2021hpdmobile} & 6 homes, 152 labelled days; 10-s clips of an 80\,Hz rectified amplitude envelope from 3 to 5 hubs per home & Home occupied, every 10\,s & CC0 & Temporal model \\
DCASE 2016 Task 2 \citep{lafay2017dcase2016} & 220 isolated office sounds (11 classes); 54 test mixtures of 120\,s & Class, onset, offset & CC BY 3.0 / 4.0 & Event sources; test set \\
DCASE 2013 OL/OS \citep{stowell2015dcase} & 320 isolated office sounds (16 classes); 11 scripted recordings in real offices (1,601\,s) & Class, onset, offset (two annotators) & CC BY 4.0 & Event sources; real test set \\
LibriSpeech \citep{panayotov2015librispeech} & train-clean-100 (251 speakers), dev-clean (40), test-clean (40) & Transcript, speaker, sex & CC BY 4.0 & Speech sources; attacks \\
BUT ReverbDB \citep{szoke2019but} & 9 rooms (3 offices, 2 lecture rooms, meeting room, conference room, hotel room, stairs), 31 microphones; empty-room noise (60\,s) and impulse responses (1\,s) & None & CC BY 4.0 & Backgrounds; room acoustics \\
\bottomrule
\end{tabular}
\end{table}

\subsection{Office benchmark}
\label{sec:office}
Training examples are 10-s scenes synthesized on the GPU. Each scene starts from a random 10-s excerpt of one empty-room recording of one training room. We add 0 to 4 isolated sounds at random positions, with event-to-background ratios drawn from $-10$ to $15$\,dB. With probability 0.4 we also add a 2 to 8\,s excerpt of one utterance at a signal-to-noise ratio from $-5$ to $20$\,dB. Every source is convolved with an impulse response of the same room, drawn independently for each source, so the sources of a scene can sit at different positions; a random gain of $\pm10$\,dB is then applied to the mixture. A second is labelled with a class if any 10-ms frame of the dry source lies within 30\,dB of the source's loudest frame. From each room we keep a random subset of at most 300 impulse responses and 24 empty-room recordings across all microphone positions and loudspeaker setups; the empty-room recordings of one BUT setup, in which a radio played in the background, are excluded.

Training uses six rooms: two offices, stairs, a hotel room, a conference room and one lecture room. It also uses, from each DCASE dataset, 15 of the 20 isolated recordings of each class, the first 15 in file-name order, and 24 utterances from each of the 251 train-clean-100 speakers. The 11 classes of DCASE 2016 are a subset of the 16 classes of DCASE 2013 (its keys-drop class is treated as keys), so both collections feed one 16-class label set. We test on three sets. The first has 1,200 synthetic scenes (12,000\,s) from three unseen rooms: office L212, lecture room D105 and meeting room C236. These scenes use the remaining 5 recordings per class and test-clean speech. The second is the 11 real office recordings of DCASE 2013 OL, labelled with the union of both annotators. The third is the 54 DCASE 2016 test mixtures (6,480\,s). Human-generated sounds are present in 44.7\%, 41.7\% and 39.4\% of the seconds in these sets. Each configuration is trained once, for a fixed 3,000 steps of 48 scenes (about 400 hours of audio), with AdamW (weight decay $10^{-4}$), gradient clipping at norm 5 and a one-cycle learning rate up to $2\times10^{-3}$; there is no validation set and no early stopping.

Labels in all three office test sets mark sound events, not presence, so a quiet second is negative even if someone is in the room. The synthetic test set holds out rooms, individual sound recordings and speakers, but its isolated sounds come from the same DCASE collections as the training sounds and may share sessions, performers and equipment, so it measures generalization to new rooms and voices more than to new recording setups. The same caveat applies to the DCASE 2013 OL recordings, which come from one of those challenges. Whether the pool of 40 sounds per class behind the DCASE 2016 test mixtures overlaps the 20 released training sounds is not documented, so we checked directly. For each of the 1,620 annotated events in the test mixtures, we computed the maximum normalized cross-correlation between the mixture segment and every released training sound of the same class. The median was 0.20 and no event exceeded 0.69, whereas for the development mixtures, which the organizers built from the training sounds, the median was 0.78 and 63\% of the events exceeded 0.7. The test mixtures therefore do not contain the training sounds.

\subsection{Baselines}
We compare the encoder with three fixed representations. Log-mel frames at 100\,Hz (64 values per second) serve as a high-information reference for the privacy attacks. Per-second log-mel statistics take the mean and standard deviation of each band (128 values per second), with and without NF. The envelope follows the HPDmobile processing (8\,kHz audio, mean shift, full-wave rectification, every 100th sample kept), giving 80 values per second. For fixed representations, the same activity head is trained on the same scenes.

\subsection{Metrics}
\label{sec:metrics}
In the office sets, each second is scored with the highest probability the activity head assigns to any human sound class, and that score is compared with the label of the same second. In the homes, each 10-s slot is scored with the predicted occupancy probability and compared with the label of the same slot. Neither score is smoothed over time. We report the area under the ROC curve (AUC), which is threshold-free and equals the probability that a random positive second scores above a random negative one (0.5 is chance), and the F1 score at a fixed threshold of 0.5, which also depends on how well the scores are calibrated. Appendix~\ref{app:nonspeech} adds average precision (AP) and the equal error rate (EER), the error at the threshold where the false positive and false negative rates are equal. For homes we add balanced accuracy, the mean recall over the two classes.

\subsection{Privacy attacks}
\label{sec:attacks}
All attacks use LibriSpeech speakers that the encoder never heard.
\begin{itemize}
\item \textbf{Speaker identification:} the utterances of the 40 test-clean speakers are split alternately into 1,319 for training the attacker and 1,301 for testing. The attacker is a two-layer bidirectional gated recurrent unit (GRU) network with 128 units per direction over the standardized representation sequence, with mean and max pooling over time, trained for 40 epochs with AdamW (learning rate $10^{-3}$, weight decay $10^{-2}$, batches of 32 utterances); for 100\,Hz inputs, strided convolutions first reduce the rate to 25\,Hz. Chance is 2.5\%.
\item \textbf{Sex classification:} the same architecture is trained for 20 epochs on dev-clean (2,703 utterances, 40 speakers) and tested on test-clean (2,620 utterances, 40 other speakers). The majority class is 53.0\%.
\item \textbf{Speech reconstruction:} a decoder (linear layer, convolution, two-layer bidirectional GRU, 256 units) maps the representation, repeated to 100\,Hz, to an 80-band log-mel spectrogram. It is trained for 15 epochs on dev-clean with the same learning rate and weight decay, and tested on 200 test-clean utterances of at most 30\,s. Each reconstruction is inverted with 60 iterations of Griffin-Lim \citep{griffin1984signal} and transcribed with Whisper small.en \citep{radford2023whisper}. We report the short-time objective intelligibility (STOI) \citep{taal2011stoi}, which is higher for more intelligible speech, and the share of reference words recovered in order,
\begin{equation}
R = \frac{\sum_u \mathrm{LCS}(w_u, \hat w_u)}{\sum_u |w_u|},
\label{eq:recall}
\end{equation}
where $w_u$ and $\hat w_u$ are the reference and recognized word sequences of utterance $u$ and LCS is the length of their longest common subsequence. Unlike the word error rate, $R$ ignores inserted words. This matters because the recognizer invents long transcripts for unintelligible input, which pushes the word error rate above 100\%.
\end{itemize}

\subsection{HPDmobile protocol}
\label{sec:hpd-protocol}
Because HPDmobile releases only the amplitude envelope, the encoder of Section~\ref{sec:method} cannot be applied to it; the temporal model is evaluated on per-slot statistics of the envelope, which stand in for the per-second embeddings of a deployment. For each home and 10-s slot, we summarize every hub's 800-sample clip by the logarithm of one plus each of six statistics: the mean, standard deviation, maximum, and 10th, 50th and 90th percentiles. We take the maximum and the mean of each feature over hubs and add the fraction of hubs reporting and a missing-data flag, giving 14 input channels. Each feature is then standardized per home with that home's median and interquartile range, without labels. Days without data are kept as gaps so that the time axis is continuous; slots without data enter the network as zeros, marked by the flag, and training windows in which more than half of the slots have no data are skipped. We evaluate leave-one-home-out, training on five homes and testing on the sixth, for every home. Pretraining uses either the five training homes (strict) or all six homes without labels (transductive), since a deployed sensor accumulates unlabelled data at its own site. The baselines are class-balanced logistic regression and gradient boosting (scikit-learn defaults otherwise) on hand-crafted features: the 14 inputs above, the time of day and trailing means over 1, 5, 30 and 120 minutes of four of the inputs (the clip mean and clip maximum, each as maximum and as mean over hubs); like the probe, they are fitted on every sixth labelled slot. We report balanced accuracy, F1 of the occupied class at a threshold of 0.5 and AUC, averaged over the six held-out homes. For full-label runs, we average over three seeds. For label efficiency we keep the labels of 1, 3 or 7 random days per training home.

\section{Results}
\label{sec:results}

\subsection{Detecting human-generated sounds}
\label{sec:utility}
Table~\ref{tab:utility} reports per-second detection of human-generated sounds for the encoder with and without NF, with speaker-adversarial training ($\lambda=1$ and 3; without NF only $\lambda=1$) and with an 8-dimensional embedding, against three fixed representations. To judge the differences, we resampled whole test recordings with replacement (2,000 bootstrap draws) and computed 95\% intervals. The differences quoted below are means of the paired differences over these draws, so they can deviate in the last digit from a subtraction of the table entries. Appendix~\ref{app:ci} lists the intervals together with results for each synthetic test room. The learned encoder outperforms both fixed representations on all three test sets. The gap is largest on real office recordings, where the encoder with NF exceeds log-mel statistics with NF by 0.109 (95\% interval 0.067 to 0.147). NF adds 0.035 (0.010 to 0.059) to the encoder on real offices and doubles its F1 at a fixed threshold, from 0.25 to 0.50. This improvement suggests that noise-floor normalization reduces sensitivity to differences in background acoustic conditions across rooms and microphones.

Adversarial training changes the real-office AUC by $-0.007$ ($-0.019$ to 0.003) for $\lambda=1$ and by 0.005 ($-0.012$ to 0.023) for $\lambda=3$. Its effect on AUC is therefore within the interval noise; the F1 at the fixed threshold does fall, so the scores need recalibration (Section~\ref{sec:limitations}). The 8-dimensional embedding loses 0.010 (0.002 to 0.020) on real offices, and adversarial training at this size loses another 0.120. Across the three synthetic test rooms, the AUC of the encoder with NF ranges from 0.939 to 0.963. The room thus matters more than the choice among the better configurations.

\paragraph{Sounds other than speech.} To evaluate detection of non-speech human-generated sounds, Appendix~\ref{app:nonspeech} excludes segments containing speech and treats other human-generated sound events as positive examples. In the real office recordings, where speech covers only 68 of the 668 positive seconds, the picture is unchanged: the encoder with NF reaches an AUC of 0.862 (0.839 to 0.884), 0.117 (0.074 to 0.155) above log-mel statistics with NF, and the adversarial variants stay within 0.01 of it. On the synthetic sets the non-speech AUC is lower than the pooled AUC (0.908 against 0.946 in unseen rooms for the encoder with NF), because speech is the easiest class to detect and the non-speech sounds are short and often quiet. Average precision and equal error rate show broadly similar trends, although the ranking of individual configurations differs across metrics.

\begin{table}[t]
\centering\small
\caption{Per-second detection of human-generated sounds on the three office test sets of Section~\ref{sec:office}: AUC and F1 at a threshold of 0.5. NF: noise-floor normalization; adv.\ $\lambda$: speaker-adversarial training with weight $\lambda$; $d$: embedding size (32 unless stated).}
\label{tab:utility}
\begin{tabular}{@{}lcccccc@{}}
\toprule
& \multicolumn{2}{c}{Unseen rooms (synthetic)} & \multicolumn{2}{c}{Real offices (DCASE 2013)} & \multicolumn{2}{c}{DCASE 2016 test} \\
\cmidrule(lr){2-3}\cmidrule(lr){4-5}\cmidrule(lr){6-7}
Representation & AUC & F1 & AUC & F1 & AUC & F1 \\
\midrule
HPDmobile-style envelope & 0.699 & 0.400 & 0.621 & 0.275 & 0.726 & 0.263 \\
Log-mel statistics & 0.873 & 0.684 & 0.682 & 0.041 & 0.609 & 0.159 \\
Log-mel statistics + NF & 0.894 & 0.699 & 0.760 & 0.447 & 0.895 & 0.460 \\
Encoder & 0.959 & 0.852 & 0.834 & 0.251 & 0.950 & 0.436 \\
Encoder + adv.\ 1 & 0.958 & 0.751 & 0.832 & 0.170 & 0.935 & 0.365 \\
Encoder + NF & 0.947 & 0.837 & 0.869 & 0.502 & 0.961 & 0.753 \\
Encoder + NF + adv.\ 1 & 0.949 & 0.779 & 0.863 & 0.354 & 0.957 & 0.609 \\
Encoder + NF + adv.\ 3 & 0.945 & 0.698 & 0.874 & 0.237 & 0.964 & 0.354 \\
Encoder + NF, $d=8$ & 0.954 & 0.830 & 0.859 & 0.462 & 0.966 & 0.705 \\
Encoder + NF + adv.\ 3, $d=8$ & 0.913 & 0.687 & 0.740 & 0.315 & 0.928 & 0.425 \\
\bottomrule
\end{tabular}
\end{table}

\subsection{What an attacker can recover}
\label{sec:privacy}
Frame-level log-mel features retain substantial speech and speaker information, with 94.3\% of reference words recovered, the same as Griffin-Lim inversion of the true spectrogram and close to the 94.5\% that Whisper recovers from the clean audio, and 98.2\% speaker-identification accuracy (Table~\ref{tab:privacy}). Reducing the representation rate to one vector per second substantially limits speech-content recovery under the tested attack. At most 1.0\% of the reference words are recovered, and STOI falls to 0.37 to 0.44. This shows that our decoder and recognizer fail, not that every attacker would.

Per-second log-mel statistics, however, still identify 97.8\% of utterances among 40 speakers, so the speaker survives the lower rate. A likely explanation is that speaker-specific characteristics, such as pitch range and timbre, remain relatively stable over an utterance and survive one-second averaging, unlike the rapidly changing acoustic information needed to recover words. Classical speaker recognition rests on the same fact \citep{reynolds2000gmm}. The task-trained encoder reduces speaker-identification accuracy to 76.8\%, which falls further to 68.6\% with NF and to 60.3\% ($\lambda=1$) and 55.3\% ($\lambda=3$) with adversarial training. Sex classification falls from 75.3\% to 60.1\%, close to the 53.0\% majority. The HPDmobile-style envelope reveals the least identity (22.4\%) but also detects human-generated sounds worst.

Shrinking the embedding lowered identification more than adversarial training did. With NF and no adversary, the 8-dimensional embedding identifies 46.6\% of utterances, fewer than any 32-dimensional encoder. Adversarial training at this size lowers identification to 33.6\%, but it costs detection and raises sex classification from 66.7\% to 85.5\%. The adversary is trained to remove speaker identity, not sex, so lowering one attribute need not lower the other.

Sending only the class probabilities lowered identification further. In this setting the device transmits the 16 outputs of the activity head instead of the embedding; since the server computes the human-sound score of Eq.~\eqref{eq:score} from these outputs anyway, detection is unchanged. The attacker identified 47.1\% of utterances from the probabilities of the encoder with NF. With adversarial training ($\lambda=3$), this fell to 37.5\%, the lowest value among the settings that keep detection performance. Sex classification reached 65.4\% and 60.3\%. This costs flexibility: the server sees only the 16 trained classes, and adding a class requires updating the device.

\begin{figure}[t]
\centering
\includegraphics[width=0.62\linewidth]{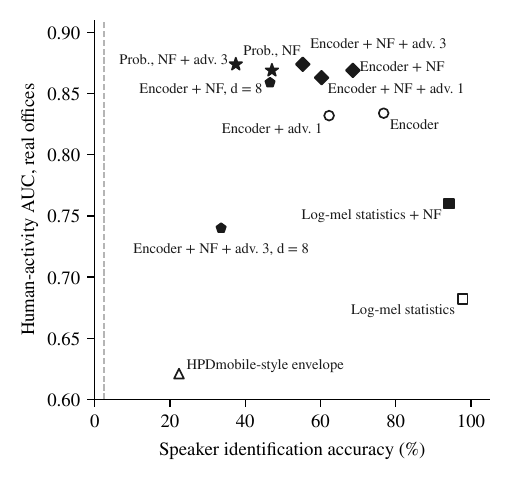}
\caption{Speaker identification accuracy among 40 speakers (Table~\ref{tab:privacy}) against the human-generated sound detection AUC on the real office recordings (Table~\ref{tab:utility}), one marker per representation. Filled markers use noise-floor normalization (NF); pentagons are 8-dimensional embeddings; stars (Prob.) send only the 16 class probabilities. The dashed line marks chance identification (2.5\%).}
\label{fig:tradeoff}
\end{figure}

Several configurations retain a similar real-office detection AUC while lowering speaker identification, although identification remains well above chance, at 38 to 55\% of the utterances (Figure~\ref{fig:tradeoff}). The test favors the attacker, since the speech is clean and close-talking, every candidate is known and the attacker trains on labelled speech of each one. A real deployment would capture far-field speech mixed with room noise, which we have not tested. Until such conditions are measured, the transmitted data should be treated as able to identify speakers, whatever the word recovery.

\begin{table}[t]
\centering\small
\caption{Privacy attacks on LibriSpeech test-clean speakers, whom the encoder never heard (Section~\ref{sec:attacks}). Rate $\times$ size: vectors per second and their dimension. Speaker ID: closed-set identification among 40 speakers, chance 2.5\%; Sex: majority class 53.0\%. Probabilities: the device sends the 16 class probabilities instead of the embedding.}
\label{tab:privacy}
\begin{tabular}{@{}llcccc@{}}
\toprule
Representation & Rate $\times$ size & Words (\%) & STOI & Speaker ID (\%) & Sex (\%) \\
\midrule
Log-mel frames (reference) & 100\,Hz $\times$ 64 & 94.3 & 0.96 & 98.2 & 94.3 \\
Log-mel statistics & 1\,Hz $\times$ 128 & 0.8 & 0.39 & 97.8 & 88.6 \\
Log-mel statistics + NF & 1\,Hz $\times$ 128 & 1.0 & 0.39 & 94.2 & 80.5 \\
HPDmobile-style envelope & 1\,Hz $\times$ 80 & 0.9 & 0.44 & 22.4 & 55.8 \\
Encoder & 1\,Hz $\times$ 32 & 0.9 & 0.39 & 76.8 & 81.9 \\
Encoder + adv.\ 1 & 1\,Hz $\times$ 32 & 0.4 & 0.39 & 62.3 & 82.6 \\
Encoder + NF & 1\,Hz $\times$ 32 & 0.7 & 0.39 & 68.6 & 75.3 \\
Encoder + NF + adv.\ 1 & 1\,Hz $\times$ 32 & 0.5 & 0.38 & 60.3 & 73.2 \\
Encoder + NF + adv.\ 3 & 1\,Hz $\times$ 32 & 0.5 & 0.39 & 55.3 & 60.1 \\
Encoder + NF & 1\,Hz $\times$ 8 & 0.1 & 0.38 & 46.6 & 66.7 \\
Encoder + NF + adv.\ 3 & 1\,Hz $\times$ 8 & 0.2 & 0.39 & 33.6 & 85.5 \\
Probabilities, Encoder + NF & 1\,Hz $\times$ 16 & 0.3 & 0.38 & 47.1 & 65.4 \\
Probabilities, Encoder + NF + adv.\ 3 & 1\,Hz $\times$ 16 & 0.1 & 0.37 & 37.5 & 60.3 \\
\bottomrule
\end{tabular}
\end{table}

\subsection{Occupancy in homes}
\label{sec:homes}
\begin{figure}[t]
\centering
\includegraphics[width=0.55\linewidth]{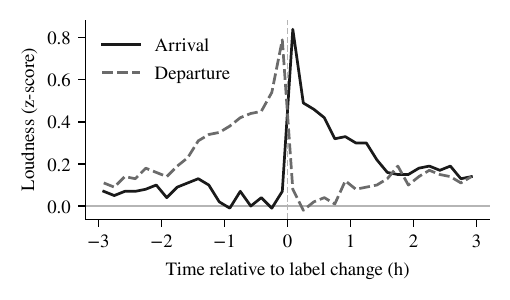}
\caption{Loudness around changes of the occupancy label in HPDmobile: z-score of the per-minute maximum envelope in 10-min bins, averaged over the six homes, from 3\,h before to 3\,h after an arrival (solid) or a departure (dashed). Time 0 is the label change.}
\label{fig:transitions}
\end{figure}

\paragraph{Label alignment and the role of transitions.} HPDmobile labels mark whether anyone is home, nights included, and loudness alone predicts this poorly: taking the log mean envelope of each slot, its maximum over hubs and a 30-minute moving average, the AUC against the labels is 0.70, 0.52, 0.67, 0.44, 0.60 and 0.39 for the six homes, below chance in two of them. Shifting the labels against this loudness by $-12$ to $+12$\,h in steps of 30\,min did not reveal a clock offset. Averaging the per-minute loudness around label changes instead shows that every home is loudest in the 10 minutes after an arrival and before a departure (Figure~\ref{fig:transitions}; peak z-scores 0.63 to 1.01). Occupancy is thus carried by sparse transition events and must be remembered through the quiet hours between them.

\paragraph{Leave-one-home-out results.} Table~\ref{tab:hpd} reports the mean over the six held-out homes, averaged over three seeds. The linear probe on the pretrained temporal encoder matches, but does not beat, the best hand-crafted baselines in AUC (0.752 under both strict and transductive pretraining, against 0.748 and 0.743), although their longest trailing mean spans only 120 minutes; the results therefore do not show that context beyond two hours was needed, and we have not varied the receptive field. Its balanced accuracy is the highest (0.678 and 0.673, against 0.649 and 0.654). Pretraining on the unlabelled audio of the test home did not change these results beyond the variation between seeds. Training the same network end to end, with or without pretraining, generalizes worse to unseen homes (AUC 0.68 to 0.71), which suggests overfitting to the five training homes.

\begin{table}[t]
\centering\small
\caption{Home occupancy on HPDmobile, leave-one-home-out (Section~\ref{sec:hpd-protocol}), averaged over the six held-out homes and three seeds; subscripts give the standard deviation over seeds. Strict: pretraining on the five training homes; transductive: also on the unlabelled audio of the test home. The hand-crafted baselines use no pretraining and gave the same result for every seed.}
\label{tab:hpd}
\begin{tabular}{@{}lcccccc@{}}
\toprule
& \multicolumn{3}{c}{Strict} & \multicolumn{3}{c}{Transductive} \\
\cmidrule(lr){2-4}\cmidrule(lr){5-7}
Method & Bal.\ acc. & F1 & AUC & Bal.\ acc. & F1 & AUC \\
\midrule
Hand-crafted + logistic regression & 0.649 & 0.676 & 0.748 & 0.649 & 0.676 & 0.748 \\
Hand-crafted + gradient boosting & 0.654 & 0.712 & 0.743 & 0.654 & 0.712 & 0.743 \\
Same network, labels only & 0.614\sd{.007} & 0.713\sd{.032} & 0.708\sd{.008} & 0.608\sd{.028} & 0.706\sd{.037} & 0.699\sd{.007} \\
Pretrained + fine-tuning & 0.605\sd{.018} & 0.704\sd{.042} & 0.684\sd{.033} & 0.596\sd{.018} & 0.658\sd{.032} & 0.686\sd{.020} \\
Pretrained + linear probe & 0.678\sd{.006} & 0.716\sd{.010} & 0.752\sd{.005} & 0.673\sd{.008} & 0.697\sd{.012} & 0.752\sd{.003} \\
\bottomrule
\end{tabular}
\end{table}

\begin{figure}[t]
\centering
\includegraphics[width=0.55\linewidth]{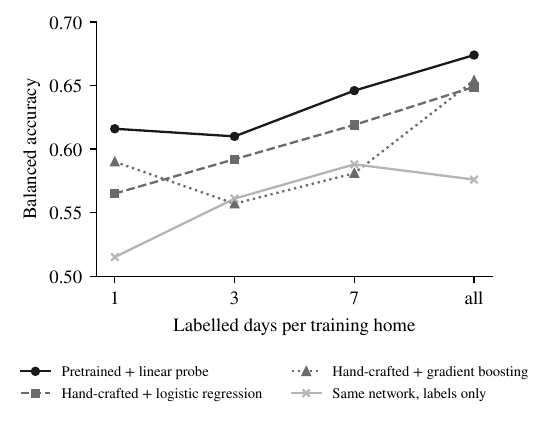}
\caption{Label efficiency on HPDmobile: balanced accuracy, averaged over the six held-out homes, when each training home keeps the labels of 1, 3 or 7 random days or of all days (transductive pretraining, seed 0).}
\label{fig:labels}
\end{figure}

\paragraph{Label efficiency.} With fewer labels, the advantage of pretraining grows (Figure~\ref{fig:labels}, one seed). With one labelled day per training home, the probe reaches a balanced accuracy of 0.616, against 0.565 and 0.590 for the hand-crafted baselines. The network trained from scratch reaches only 0.515. In AUC, the probe stays best or tied at 1, 3 and 7 labelled days (0.691, 0.706 and 0.722).

\section{Limitations}
\label{sec:limitations}

\paragraph{What was measured.} The office experiments evaluate per-second human-generated sound detection rather than occupancy, because the test sets provide sound-event labels but no occupancy labels. Therefore, the non-speech evaluation in Appendix~\ref{app:nonspeech} demonstrates the detection of audible human-generated sounds, not the presence of silent occupants. The training scenes are synthetic, and the only real office test set is small (11 scripted recordings, 1,601\,s). It also comes from the same challenge as part of our training sounds. Each office configuration was trained once, so our intervals cover the variation between test recordings but not between training runs. We did not compare the encoder with audio taggers pretrained on large labelled collections, such as PANNs \citep{kong2020panns}, which would show how far training on synthetic scenes is from the detection that larger models reach. The temporal model was trained on homes, with whole-home labels and an amplitude envelope as input. The two stages of Figure~\ref{fig:overview} have therefore not yet been evaluated together. Its advantage over hand-crafted features was small with all labels and larger with one to seven labelled days, which we ran with a single seed. Deployability was argued from the size of the encoder (0.59 million parameters); we have not measured its latency or power on a target device.

\paragraph{Privacy.} Speaker identity was reduced but not removed, and under adversarial training sex classification can move in the opposite direction from identity (Section~\ref{sec:privacy}). The adversary used during training, a perceptron over pooled embeddings, is weaker than the recurrent attacker we evaluate, so adversarial training may understate what a matched adversary could remove. The attacks also cover only part of the threat: a closed set of 40 speakers, one language and attackers that we trained ourselves, so a stronger attacker could recover more. Closed-set identification is also a different task from the open-set verification with which speaker anonymization is evaluated \citep{tomashenko2022voiceprivacy}, and a drop in closed-set accuracy need not carry over to verification \citep{srivastava2019privacy}, so our numbers are not comparable with that literature.

\paragraph{Calibration.} On real office recordings, the F1 at a fixed threshold ranges from 0.04 to 0.50 across representations and falls with stronger adversarial training, even when the AUC does not. Scores therefore have to be calibrated for each room and model before a threshold can drive control.

\section{Future Work}
\label{sec:future}

The main next step is an office dataset of our own, to be reported in a separate journal article. We plan to record single-occupant labs, shared labs and lecture rooms on the University of Illinois campus, several rooms of each kind so that whole rooms can be held out, each for at least two weeks including nights and weekends. The same low-cost device, a single-board computer with a small digital microphone, will sit next to a commercial PIR sensor in every room, and ground truth will come from sign-in sheets, door sensors and, in lecture rooms, class schedules. Sessions in which participants sit quietly, for example reading for 30 minutes, will test the case that PIR sensors miss, and for both sensors we will measure how often occupied periods are judged vacant and how long each takes to switch off after people leave. Training the temporal model on these office streams will evaluate the two stages together. Only the representations computed on the device will be stored long term; raw audio will be recorded only in short sessions with consenting participants, and the study will begin after approval by the university's Institutional Review Board. On the public benchmarks, the next steps are to repeat the label-efficiency study and the office configurations over several seeds, to pretrain across many buildings, to add a pretrained audio tagger as a reference for detection, to train the adversary with the attacker's architecture, to vary the receptive field of the temporal model, and to test simple temporal smoothing of the per-second scores.

For privacy, we will add quantization and noise to the transmitted class probabilities and train an adversary for sex next to the one for speaker identity. The attacks will be repeated with far-field speech recorded in the target rooms, with open-set speaker verification and with other languages. For calibration, we will use empty periods such as nights, which occur at every site and need no labels. The threshold of each room can then be set from the scores observed during those periods. Beyond this paper, we plan to fuse the audio representations with PIR and LiDAR sensing. We also plan to evaluate closed-loop lighting and ventilation control for energy and comfort, and to extend the study to homes.

\section{Conclusion}
A one-second acoustic representation computed on the device can detect the sounds people make while working, such as typing, page turns and coughs, in rooms it has never heard. Our reconstruction attack recovered almost none of the words from it. Who is speaking, however, survives the reduction in rate and has to be removed deliberately. Noise-floor normalization combined with adversarial training, a smaller embedding or class-probability outputs lowered speaker identification from 77\% to 38 to 55\%. Detection stayed at a real-office AUC of 0.86 to 0.87, but identification is still far from chance. In homes, occupancy shows up as short loud transitions separated by long quiet stretches. A temporal model pretrained without labels makes the best use of this structure when labels are scarce. Office recordings with occupancy labels are the next step.

\section*{Data and Code Availability}
All datasets are public (Table~\ref{tab:data}). Code for data preparation, training, attacks, the bootstrap analysis, the non-speech evaluation and the DCASE 2016 overlap check is available at \url{https://github.com/AaronChae/occupancy-audio-privacy}.

\section*{Ethics Statement}
This study uses only public datasets under their stated licenses and collects no new data from people. The planned office recordings will follow the Institutional Review Board process described in Section~\ref{sec:future}.

\section*{Acknowledgments}
This research used the Delta advanced computing and data resource which is supported by the National Science Foundation (award OAC 2005572) and the State of Illinois. Delta is a joint effort of the University of Illinois Urbana-Champaign and its National Center for Supercomputing Applications.

\bibliographystyle{plainnat}
\bibliography{references}

\appendix
\section{Per-Home Results}
\begin{table}[h]
\centering\small
\caption{AUC on each held-out home of HPDmobile (strict pretraining, seed 0). The last two rows give the number of labelled days and the share of occupied 10-s slots in each home.}
\begin{tabular}{@{}lcccccc@{}}
\toprule
Method & H1 & H2 & H3 & H4 & H5 & H6 \\
\midrule
Hand-crafted + logistic regression & 0.861 & 0.698 & 0.833 & 0.717 & 0.760 & 0.620 \\
Hand-crafted + gradient boosting & 0.700 & 0.745 & 0.889 & 0.702 & 0.806 & 0.616 \\
Same network, labels only & 0.622 & 0.706 & 0.871 & 0.607 & 0.748 & 0.639 \\
Pretrained + fine-tuning & 0.671 & 0.578 & 0.845 & 0.573 & 0.795 & 0.555 \\
Pretrained + linear probe & 0.826 & 0.708 & 0.815 & 0.707 & 0.761 & 0.666 \\
\midrule
Labelled days & 30 & 17 & 35 & 17 & 15 & 38 \\
Occupied share (\%) & 82.2 & 60.4 & 78.2 & 81.7 & 75.6 & 46.8 \\
\bottomrule
\end{tabular}
\end{table}

\section{Bootstrap Intervals and Per-Room Results}
\label{app:ci}
Table~\ref{tab:ci} gives 95\% percentile intervals of the AUC for human-generated sounds, obtained by resampling whole recordings with replacement (2,000 draws): 11 recordings for real offices and 54 mixtures for DCASE 2016. The last three columns give the AUC in each unseen synthetic room, computed on 400 new scenes per room, so they differ slightly from the pooled synthetic results in Table~\ref{tab:utility}. Paired differences between configurations, computed on the same draws, are reported in the code repository.

\begin{table}[h]
\centering\small
\caption{AUC for human-generated sounds: 95\% bootstrap intervals on the two recorded test sets, and AUC in each of the three unseen synthetic rooms.}
\label{tab:ci}
\begin{tabular}{@{}lccccc@{}}
\toprule
Representation & Real offices & DCASE 2016 test & L212 & D105 & C236 \\
\midrule
HPDmobile-style envelope & 0.621 (0.587--0.655) & 0.726 (0.695--0.757) & 0.704 & 0.694 & 0.748 \\
Log-mel statistics & 0.682 (0.617--0.740) & 0.609 (0.587--0.631) & 0.869 & 0.855 & 0.879 \\
Log-mel statistics + NF & 0.760 (0.726--0.794) & 0.895 (0.884--0.905) & 0.907 & 0.913 & 0.885 \\
Encoder & 0.834 (0.816--0.854) & 0.950 (0.942--0.957) & 0.945 & 0.959 & 0.953 \\
Encoder + adv.\ 1 & 0.832 (0.817--0.847) & 0.935 (0.927--0.943) & 0.942 & 0.958 & 0.946 \\
Encoder + NF & 0.869 (0.847--0.892) & 0.961 (0.954--0.967) & 0.948 & 0.963 & 0.939 \\
Encoder + NF + adv.\ 1 & 0.863 (0.833--0.888) & 0.957 (0.950--0.963) & 0.948 & 0.964 & 0.930 \\
Encoder + NF + adv.\ 3 & 0.874 (0.862--0.887) & 0.964 (0.958--0.970) & 0.944 & 0.960 & 0.929 \\
Encoder + NF, $d=8$ & 0.859 (0.831--0.884) & 0.966 (0.961--0.972) & 0.949 & 0.965 & 0.948 \\
Encoder + NF + adv.\ 3, $d=8$ & 0.740 (0.699--0.774) & 0.928 (0.920--0.936) & 0.919 & 0.930 & 0.885 \\
\bottomrule
\end{tabular}
\end{table}

\section{Detection of Non-Speech Human Sounds}
\label{app:nonspeech}
Table~\ref{tab:nonspeech} restricts the evaluation to the seconds whose labels contain no speech. A second is positive if a non-speech human sound (clear throat, cough, door slam, drawer, keyboard, keys, knock, laughter, mouse click, page turn, pen drop or light switch) is active and negative if no human sound is active; seconds with speech are dropped. This leaves 600 positive and 933 negative seconds in the real offices, 2,127 and 3,930 in the DCASE 2016 test mixtures, and 2,760 and 6,624 in the 1,200 synthetic scenes of Appendix~\ref{app:ci}. The score is the same human-sound score as in Table~\ref{tab:utility}; intervals are 95\% bootstrap intervals over whole recordings (2,000 draws), as in Table~\ref{tab:ci}. Per-second results for all seconds, with AP and EER, are in the code repository.

\begin{table}[h]
\centering\footnotesize
\caption{Detection of non-speech human sounds on the seconds without speech: AUC with 95\% bootstrap interval, average precision (AP) and equal error rate (EER).}
\label{tab:nonspeech}
\setlength{\tabcolsep}{2.6pt}
\begin{tabular}{@{}lccccccccc@{}}
\toprule
& \multicolumn{3}{c}{Real offices (DCASE 2013)} & \multicolumn{3}{c}{DCASE 2016 test} & \multicolumn{3}{c}{Unseen rooms (synthetic)} \\
\cmidrule(lr){2-4}\cmidrule(lr){5-7}\cmidrule(lr){8-10}
Representation & AUC & AP & EER & AUC & AP & EER & AUC & AP & EER \\
\midrule
HPDmobile-style envelope & 0.622 (0.589--0.655) & 0.564 & 0.418 & 0.713 (0.683--0.742) & 0.657 & 0.352 & 0.683 & 0.467 & 0.363 \\
Log-mel statistics & 0.660 (0.592--0.723) & 0.593 & 0.395 & 0.542 (0.519--0.566) & 0.545 & 0.479 & 0.775 & 0.598 & 0.280 \\
Log-mel statistics + NF & 0.745 (0.709--0.782) & 0.673 & 0.312 & 0.879 (0.868--0.890) & 0.841 & 0.173 & 0.832 & 0.654 & 0.214 \\
Encoder & 0.825 (0.803--0.846) & 0.804 & 0.248 & 0.941 (0.931--0.949) & 0.932 & 0.106 & 0.915 & 0.816 & 0.143 \\
Encoder + adv.\ 1 & 0.824 (0.807--0.838) & 0.786 & 0.260 & 0.924 (0.915--0.933) & 0.894 & 0.120 & 0.911 & 0.775 & 0.147 \\
Encoder + NF & 0.862 (0.839--0.884) & 0.839 & 0.220 & 0.953 (0.946--0.960) & 0.941 & 0.089 & 0.908 & 0.771 & 0.140 \\
Encoder + NF + adv.\ 1 & 0.855 (0.826--0.881) & 0.822 & 0.217 & 0.949 (0.941--0.957) & 0.936 & 0.100 & 0.905 & 0.748 & 0.143 \\
Encoder + NF + adv.\ 3 & 0.869 (0.857--0.882) & 0.817 & 0.200 & 0.958 (0.950--0.965) & 0.942 & 0.084 & 0.899 & 0.734 & 0.152 \\
Encoder + NF, $d=8$ & 0.850 (0.823--0.874) & 0.829 & 0.236 & 0.960 (0.954--0.966) & 0.947 & 0.089 & 0.915 & 0.784 & 0.134 \\
Encoder + NF + adv.\ 3, $d=8$ & 0.719 (0.679--0.752) & 0.595 & 0.308 & 0.916 (0.907--0.925) & 0.826 & 0.102 & 0.845 & 0.638 & 0.194 \\
\bottomrule
\end{tabular}
\end{table}

\end{document}